\documentclass[twocolumn,english,reprint,amsmath,amssymb,aps]{revtex4-1}
\usepackage[T1]{fontenc}
\usepackage[latin9]{inputenc}
\usepackage{verbatim}
\usepackage{esint}
\usepackage{color}

\makeatletter
\usepackage{graphicx}
\usepackage{dcolumn}
\usepackage{bm}

\makeatother

\usepackage{babel}
\begin{document}

\title{ Two non-perturbative  $\alpha^{\prime}$ corrected or loop  corrected  string cosmological solutions}

\author{Li Song}
\email{songli1984@scu.edu.cn}
\affiliation{College of Physics, Sichuan University, Chengdu, 610065, China}

\author{Deyou Chen}
\email{deyouchen@hotmail.com}
\affiliation{School of Science, Xihua University, Chengdu 610039, China}


\begin{abstract}

\noindent In this paper, 
we present two   non-perturbative string cosmological solutions without curvature singularities for the bosonic gravi-dilaton system. These solutions are general in the sense that they can straightforwardly match the perturbative solution to arbitrarily high orders in the perturbative region. 
The first  solution includes non-perturbative $\alpha'$ corrections based on Hohm-Zwiebach action. We then use the simple phenomenological map between   $\alpha^{\prime}$ corrected theory and loop corrected theory in string cosmology to construct a non-perturbative loop corrected non-singular solution. Both   solutions are non-singular everywhere. Therefore, the pre- and post-big-bangs are smoothly connected by these solutions.

\end{abstract}

\maketitle

\onecolumngrid

\section{Introduction}

Resolving the big-bang singularity is a widely concerned problem.
In Einstein's general theory of gravity, spacetime is generated from
the big-bang. On the other hand, in string cosmology, the story is
quite different due to the famous scale-factor duality \cite{Veneziano:1991ek,Sen:1991zi,Tseytlin:1991wr,Tseytlin:1991xk}.
The scale-factor duality had been firstly obtained from the equations
of motion (EOM) of the string's low energy gravi-dilaton effective
action. It shows that the EOM with the FLRW-like ansatz is invariant
under the transformations between the scale factor and its inversion,
namely $a\left(t\right)\leftrightarrow1/a\left(t\right)$. Keep
in mind that the string dilaton plays a central role in this duality.
The main differences between T-duality and scale-factor duality are
that the scale-factor duality does not require the background compactified,
and it belongs to a continuous group of $O\left(d,d\right)$. Remarkably,
the scale-factor duality introduces a pre-big-bang cosmology \cite{Gasperini:1996fu,Veneziano:2000pz,Gasperini:2002bn,Gasperini:2007vw,Gasperini:1992em}.
It implies that the story of our universe was not only born from the
initial big-bang singularity, but there exists a long evolution in
a region of pre-big-bang. Pre- and post- big-bang scenarios are disconnected
by the big-bang singularity. On the other hand, when the universe
evolves to the big-bang singularity, the growth of the string coupling
$g_{s}=\exp\left(2\phi\right)$ and the Hubble parameter $H\left(t\right)$
invalidate the perturbative theory. The theory of quantum gravity
around this region requires the low energy effective action to include
two kinds of corrections: (1) the string curvature
scale, which includes higher-derivative $\alpha^{\prime}$ corrections,
(2) the strong coupling regime, which requires the quantum loop corrections.

In literature, some phenomenological higher loop models 
\cite{Gasperini:1992em,Gasperini:2003pb,Gasperini:2004ss}
were proposed to resolve the big-bang singularity. As for the higher-derivative
$\alpha^{\prime}$ corrections, at the cost of losing the scale-factor
duality or the $O(d,d)$ symmetry, some simplified models were proposed\cite{Gasperini:1996fu, Easson:2003ia}
to smoothly connect the Pre- and post- big-bang scenarios. One may
wonder if we can obtain a nature global picture of the non-singular
universe when the action includes the $\alpha^{\prime}$ corrections
to all orders.

To answer this question, let us focus on the recent developments of
non-perturbative string cosmology to all orders in $\alpha^{\prime}$.
In refs. \cite{Meissner:1991zj,Meissner:1996sa}, in addition
to manifesting the scale-factor duality in the EOM, Meissner and Veneziano
found that when the massless closed string fields only depending on
time, the low energy effective action of closed string with the zeroth
and the first orders in $\alpha^{\prime}$ could be simplified and
rewritten in an $O\left(d,d\right)$ invariant form. After that, Sen
proved that this result can be extended to all orders in $\alpha^{\prime}$
corrections of full string field theory \cite{Sen:1991zi,Sen:1991cn}.
In other words, the closed string spacetime action can be rewritten
into the $O\left(d,d\right)$ covariant form to all orders in $\alpha^{\prime}$
without imposing any extra constraint or symmetry. Based on these
works, Hohm and Zwiebach \cite{Hohm:2015doa,Hohm:2019ccp,Hohm:2019jgu}
demonstrated that the $O\left(d,d\right)$ covariant spacetime action
can be dramatically simplified into a simple form. This result in
turn provides possible non-perturbative dS or AdS vacua to the bosonic
string theory \cite{Hohm:2019ccp,Wang:2019mwi,Krishnan:2019mkv,Nunez:2020hxx,Bieniek:2023ubx}.
Moreover, the Hohm-Zwiebach action makes it possible to analyze the
stringy effects on the cosmological singularity and black hole singularity
seriously. In ref. \cite{Wang:2019kez}, a set of non-singular non-perturbative
string cosmological solutions is constructed for the first time. This
set of solutions is subsequently extended to more general solutions
which can match the perturbative solution to arbitrary order in $\alpha$'
expansion \cite{Wang:2019dcj}. More recent developments on the smoothing  
cosmological and black hole singularities refer to refs. 
\cite{Ying:2021xse, Ying:2022xaj,Klinkhamer:2019ocj,Bernardo:2019bkz,Bernardo:2020zlc,
Quintin:2021eup,Codina:2021cxh,Ying:2022cix,Codina:2023fhy,Gasperini:2023tus}.

The purpose of this paper is to   give non-singular non-perturbative solutions with $\alpha'$ corrections or 
loop corrections. We first construct an    $\alpha'$ corrected solution based on Hohm-Zwiebach action. 
The solution we present is non-perturbative in the sense that it covers the whole region from the pre-big-bang to the post-big-bang and non-singular everywhere. In the perturbative region $t\to\infty$, this solution matches the perturbative solution to 
arbitrary order in $\alpha'$. So, though we currently only know the first two orders in $\alpha'$ corrections, once higher orders are calculated in the future, our solution can straightforwardly match them by fixing the parameters. 
Our solution is parallel to these given in \cite{Wang:2019dcj}.
In ref.  \cite{Wang:2019dcj},
a phenomenological map between the $\alpha'$ corrected theory and loop corrected theory are found. With the help of this map, we then construct a non-perturbative loop corrected solution, which is non-singular everywhere and includes higher loop contributions. 
So, both the $\alpha'$ corrected or the loop corrected solution we are going to  construct resolve the big-bang singularity and  connects the pre-big-bang
and post-big-bang scenarios smoothly. 

The reminder of this paper is outlined as follows. We give a non-singular non-perturbative solution to any order in $\alpha'$ expansion  in section 2.  In section 3, we present a loop corrected non-singular non-perturbative solution.    Section 4 is the conclusion.

\section{The general $\alpha^{\prime}$ corrected solution}

Since we are going to find both $\alpha'$ corrected solutions and loop corrected solutions, we write the full perturbative structure of the closed string effective action,
\begin{eqnarray}
I & = & \int\:d^{d+1}x\sqrt{-g}\bigg\{ e^{-2\phi}\Big[(R+4(\partial\phi)^{2}-
\frac{1}{12}{\cal H}^{2})+\frac{\text{\ensuremath{\alpha'}}}{4}(R_{\mu\nu\sigma\rho}R^{\mu\nu\sigma\rho}+\cdots)+{\cal O}(\alpha'^{2})\big]\nonumber\\
 & + & \Big[(c_{R}^{1}R+c_{\phi}^{1}(\partial\phi)^{2}+c_{{\cal H}}^{1} {\cal H}^{2})+\alpha'(c_{\alpha'R}^{1}R_{\mu\nu\sigma\rho}R^{\mu\nu\sigma\rho}+\cdots)+{\cal O}(\alpha'^{2})\Big]\nonumber\\
 & + & e^{2\phi}\Big[(c_{R}^{2}R+c_{\phi}^{2}(\partial\phi)^{2}+c_{{\cal H}}^{2} {\cal H}^{2})+\alpha'(c_{\alpha'R}^{2}R_{\mu\nu\sigma\rho}R^{\mu\nu\sigma\rho}+\cdots)+{\cal O}(\alpha'^{2})\Big]\nonumber\\
 & + & \cdots\bigg\}.
\label{eq:complete action}
\end{eqnarray}

This action contains three massless fields, the metric $g_{\mu\nu}$, dilaton $\phi$ and the  antisymmetric field $b_{\mu\nu}$ whose field strength is ${\cal H}_{\mu\nu\rho}=3\partial_{[\mu}b_{\nu\rho]}$. We will set $b_{\mu\nu}=0$ in this paper. All $c_{[\cdots]}^{i}$ are not known until now. We first consider the $\alpha'$ corrected solutions. To this end, we focus on the loop tree level, i.e. the first line of the above action. In the FLRW background, 

\begin{equation}
ds^{2}=-dt^{2}+a^{2}\left(t\right)\delta_{ij}dx^{i}dx^{j}.\label{FLRW}
\end{equation}

\noindent the loop tree level action with all  orders in $\alpha^{\prime}$ corrections is given by Hohm and
Zwiebach in Refs. \cite{Hohm:2019ccp,Hohm:2019jgu}:
\begin{eqnarray}
I_{\alpha'} & = & \int d^{D}x\sqrt{-g}e^{-2\phi}\left(R+4\left(\partial\phi\right)^{2}+\frac{1}{4}\alpha^{\prime}\left(R^{\mu\nu\rho\sigma}R_{\mu\nu\rho\sigma}+\ldots\right)+\alpha^{\prime2}(\ldots)+\ldots\right),\label{eq:original action with alpha}\\
 & = & \int dte^{-\Phi}\left(-\dot{\Phi}^{2}+\sum_{k=1}^{\infty}\left(\alpha^{\prime}\right)^{k-1}c_{k}\mathrm{tr}\left(\dot{\mathcal{S}}^{2k}\right)\right),\label{eq:Odd action with alpha}
\end{eqnarray}

\noindent The notation $\mathcal{S}$ is defined as

\noindent 
\begin{equation}
\mathcal{S}=\left(\begin{array}{cc}
0 & a^{2}\left(t\right)\\
a^{-2}\left(t\right) & 0
\end{array}\right).
\end{equation}

\noindent In the action (\ref{eq:Odd action with alpha}), we can
only determine the coefficients $c_{1}=-\frac{1}{8}$ and $c_{2}=\frac{1}{64}$
for the bosonic string theory through the one-loop and two-loops beta
functions of non-linear sigma model, and $c_{k\geq3}$ are undetermined
constants. Define

\begin{eqnarray}
H\left(t\right) & = & \frac{\dot{a}\left(t\right)}{a\left(t\right)},\nonumber \\
f\left(H\right) & = & d\sum_{k=1}^{\infty}\left(-\alpha^{\prime}\right)^{k-1}2^{2\left(k+1\right)}kc_{k}H^{2k-1}\nonumber \\
&=&-2dH-2d\alpha^{\prime}H^{3}+\mathcal{O}\left(\alpha^{\prime3}\right),\nonumber \\
g\left(H\right) & = & d\sum_{k=1}^{\infty}\left(-\alpha^{\prime}\right)^{k-1}2^{2k+1}\left(2k-1\right)c_{k}H^{2k}\nonumber \\
&=& -dH^{2}-\frac{3}{2}d\alpha^{\prime}H^{4}+\mathcal{O}\left(\alpha^{\prime}\right),\label{eq:EOM fh gh}
\end{eqnarray}

\noindent where $H\left(t\right)$ is the Hubble parameter. Note that

\[
g'(H)=Hf'(H),\quad{\rm and}\quad g(H)=Hf(H)-\int_{0}^{H}f(x)dx,
\]

\noindent where $f'(H)\equiv\frac{d}{dH}f(H)$. The action (\ref{eq:Odd action with alpha})
is simplified to the Hohm-Zwiebach action

\begin{eqnarray}
I_{HZ} & = & \int dte^{-\Phi}\left(-\dot{\Phi}^{2}+g(H)-Hf(H)\right),\label{eq:HZ action}
\end{eqnarray}

\noindent After variation, the EOM of Hohm-Zwiebach action (\ref{eq:HZ action}) is 

\begin{eqnarray}
\ddot{\Phi}+\frac{1}{2}Hf\left(H\right) & = & 0,\nonumber \\
\frac{d}{dt}\left(e^{-\Phi}f\left(H\right)\right) & = & 0,\nonumber \\
\dot{\Phi}^{2}+g\left(H\right) & = & 0.\label{eq:Original EoM}
\end{eqnarray}

\noindent In the perturbative region $t\to \infty$, the perturbative solution can be obtained iteratively by using (\ref{eq:EOM fh gh}), 

\begin{eqnarray}
H\left(t\right) & = & \frac{\sqrt{2}}{\sqrt{\alpha^{\prime}}}\left[\frac{t_{0}}{t}-160c_{2}\frac{t_{0}^{3}}{t^{3}}+\frac{256\left(770c_{2}^{2}+19c_{3}\right)}{3}\frac{t_{0}^{5}}{t^{5}}\right.\nonumber\\
 &  & \left.-\frac{2048\left(88232c_{2}^{3}+4644c_{3}c_{2}+41c_{4}\right)}{5}\frac{t_{0}^{7}}{t^{7}}+\mathcal{O}\left(\frac{t_{0}^{9}}{t^{9}}\right)\right], \nonumber\\
\Phi\left(t\right) & = & -\frac{1}{2}\log\left(\beta^2\frac{t^{2}}{t_{0}^{2}}\right)-32c_{2}\frac{t_{0}^{2}}{t^{2}}+\frac{256\left(44c_{2}^{2}+c_{3}\right)}{3}\frac{t_{0}^{4}}{t^{4}} \nonumber\\
 &  & -\frac{2048\left(6976c_{2}^{3}+352c_{3}c_{2}+3c_{4}\right)}{15}\frac{t_{0}^{6}}{t^{6}}+\mathcal{O}\left(\frac{t_{0}^{8}}{t^{8}}\right),
\label{eq:perturbative solution}
\end{eqnarray}
with

\begin{eqnarray*}
f\left(H\left(t\right)\right) & = & -2dH-128c_{2}d\alpha^{\prime}H^{3}+768c_{3}d\alpha^{\prime2}H^{5}-4096c_{4}d \alpha^{\prime3}H^{7}+\mathcal{O}\left(\alpha^{\prime4}H^{9}\right), \nonumber\\
 & = & \frac{\sqrt{d}}{t_{0}}\left[-\frac{2t_{0}}{t}+64c_{2}\frac{t_{0}^{3}}{t^{3}}-\frac{512\left(50c_{2}^{2}+c_{3}\right)}{3}\frac{t_{0}^{5}}{t^{5}}\right. \nonumber\\
 &  & \left.+\frac{4096\left(2632c_{2}^{3}+124c_{3}c_{2}+c_{4}\right)}{5} \frac{t_{0}^{7}}{t^{7}}+\mathcal{O}\left(\frac{t_{0}^{9}}{t^{9}}\right)\right],\nonumber\\
g\left(H\left(t\right)\right) & = & -dH^{2}-96c_{2}d\alpha^{\prime}H^{4} +640c_{3}d\alpha^{\prime2}H^{6}-3584c_{4}d\alpha^{\prime3} H^{8}+\mathcal{O}\left(\alpha^{\prime4}H^{10}\right),\nonumber\\
 & = & \frac{1}{t_{0}^{2}}\left[-\frac{t_{0}^{2}}{t^{2}}+128c_{2}\frac{t_{0}^{4}}{t^{4}}-\frac{2048\left(50c_{2}^{2}+c_{3}\right)}{3}\frac{t_{0}^{6}}{t^{6}}\right.\nonumber\\
 &  & \left.+\frac{8192\left(24448c_{2}^{3}+1136c_{3}c_{2}+9c_{4}\right)}{15}\frac{t_{0}^{8}}{t^{8}}+\mathcal{O}\left(\frac{t_{0}^{10}}{t^{10}}\right)\right],
\end{eqnarray*}

\noindent where $\beta$ is an integration constant, $t_{0}\equiv\frac{\sqrt{\alpha'}}{\sqrt{2d}}$ and universal $c_{1}=-\frac{1}{8}$ is used. 
One is ready to see that this solution is  singular at the non-perturbative region $t=0$.  
From the scale factor duality,  $H(t)\to-H(t)$, $\Phi(t)\to\Phi(t)$, $f(t)\to-f(t)$ and $g(t)\to g(t)$ is also a solution. 

A non-perturbative non-singular  solution of the EOM (\ref{eq:Original EoM}) should meet two conditions: (1) It must match the perturbative solution (\ref{eq:perturbative solution}) in the perturbative region $\frac{\sqrt{\alpha'}}{t}\to 0$;   (2) It must be non-singular for any  $\frac{\sqrt{\alpha'}}{t}$.  Since until now $c_{k\ge3}$ are still unknown, a good solution only needs to match the first two orders in $\alpha'$. Such a solution has been constructed in ref.   \cite{Wang:2019kez}.  However, an important question is that, can we construct a general solution which can easily match all given  $c_{k\le n}$ for some  $n>2$? In other words, a general solution 
is expressed in terms of $c_{k}$.  In ref. \cite{Wang:2019dcj}, two such solutions were given. 
In this work, we give another simpler one. Every term in our solution is non-singular. Let us first define a dimensionless parameter 
\begin{equation}
\tau \equiv \frac{t}{t_0} = \sqrt\frac{2d}{\alpha'} t.
\end{equation}
After some long calculation, we find a non-perturbative solution of the EOM (\ref{eq:Original EoM}),
\begin{eqnarray}
H(t)&=& \frac{\sqrt{\frac{2}{\alpha' \beta ^2 \lambda _0}} \left(e^{-\sum _{k=1}^{\infty } \frac{\lambda _k}{\tau ^{2 k}+1}} \left(\left(\tau ^2+1\right)^2 \sum _{k=1}^{\infty } \left(\frac{8 k^2 \lambda _k \tau ^{4 k-2}}{\left(\tau ^{2 k}+1\right)^3}-\frac{2 k (2 k-1) \lambda _k \tau ^{2 k-2}}{\left(\tau ^{2 k}+1\right)^2}\right)+\tau ^2-1\right)\right)}{\left(\tau ^2+1\right)^{3/2}}\\
\Phi(t) &=& \frac{1}{2}\log\frac{\lambda_0}{1+\tau^2} +\sum_{n=1}^\infty \frac{\lambda_n}{1+\tau^{2n}},\label{eq:alphaSolutionPhi}\\
f(H(t))&=&  -2 d \sqrt{\frac{2 \beta ^2 \lambda _0}{\left(\tau ^2+1\right) \alpha '}} e^{\sum _{k=1}^{\infty } \frac{\lambda _k}{\tau ^{2 k}+1}}  = -2d H(t) -2d\alpha' H(t)^3 + {\cal O}(\alpha^{\prime 2})  \\
g(H(t))&=& -\frac{2 d}{\alpha' } \left(\sum _{k=1}^{\infty } \frac{2 k \lambda _k \tau ^{2 k-1}}{\left(\tau ^{2 k}+1\right)^2}+\frac{\tau }{\tau ^2+1}\right)^2 =  -d H(t)^2 -\frac{3}{2} d\alpha' H(t)^4 + {\cal O}(\alpha^{\prime 2}).
\label{eq:alphaSolution}
\end{eqnarray}
Applying $c_1 =-1/8$ and $c_2=1/64$,  expanding this solution in the perturbative region $t/\sqrt{\alpha'}\to \infty$, to match the perturbative solution (\ref{eq:perturbative solution}), we identify
\begin{eqnarray}
\lambda_0 &=& 1/\beta^2,\quad \lambda_1 = 0,\quad \lambda_2= \frac{11+1024c_3}{12},\nonumber\\
\lambda_3 &=& -\frac{4}{15}(13+2816c_3+1536c_4)...
\end{eqnarray}
It is obvious that every single term in the sums such as $\sum_{n=1}^\infty \frac{\lambda_n}{1+\tau^{2n}}$  of the solution is non-singular everywhere. But we want to point out that there could be a tiny possibility that the summation is not non-singular for fine-tuned parameters $\lambda_n, n>0$ \cite{Gasperini:2023tus1}.  If choosing  $\lambda_{n\ge 1}=0$, the solution reduces to 

\begin{eqnarray}
H(t) & = & -\frac{\sqrt{2}}{\sqrt{\alpha'}}\frac{\left(1-\tau^{2}\right)}{\left(1+\tau^{2}\right)^{3/2}},  \nonumber \\
\Phi(t) & = & -\frac{1}{2}\log\beta^2-\frac{1}{2}\log\left(1+\tau^{2}\right),\nonumber\\
f(t) & = & -\frac{2\sqrt{2}d}{\sqrt{\alpha'}}\frac{1}{\sqrt{1+\tau^{2}}},\nonumber \\
g(t) & = & -\frac{2d}{\alpha'}\frac{\tau^{2}}{\left(1+\tau^{2}\right)^{2}}.
\label{eq:2nd order solution}
\end{eqnarray}
which is exactly the solution given in ref.   \cite{Wang:2019kez}.

\section{The general loop corrected solution}

After constructing the non-perturbative non-singular $\alpha'$ corrected solution, we want to find the corresponding loop corrected non-perturbative non-singular solution. We do not know the form and coefficients of the higher loop terms in the full perturbative action (\ref{eq:complete action}). For the  FLRW background (\ref{FLRW}), an effective loop corrected action  is sketched as

\begin{eqnarray}
I_{\rm Loop}&=&\int d^{d+1}x\sqrt{-g}e^{-2\phi}\left[R+4\left(\partial_{\mu}\phi\right)^{2} - V\left(e^{-\Phi\left(x\right)}\right)\right],\nonumber\\
&=& \int dt e^{-\Phi} \big[ -\dot\Phi + d H^2 -V(e^{-\Phi})\big],
\label{eq:action with potential}
\end{eqnarray}
where the $O(d,d)$ non-local dilaton is \cite{Gasperini:1992em,Gasperini:2003pb},
\begin{equation}
e^{-\Phi\left(t\right)}=V_{d}\int dt'\left|\frac{d\left(2\phi\right)}{dt'} \right|\sqrt{-g\left(t'\right)}e^{-2\phi\left(t'\right)}\delta\left(2\phi\left(t\right)-2\phi\left(t^{\prime}\right)\right)=V_{d}\sqrt{-g\left(t\right)}e^{-2\phi\left(t\right)}.
\label{eq:reduced non-local dilaton}
\end{equation}
The EOM is  
\begin{eqnarray}
2\ddot{\Phi}_L -2 dH_L^2 - \frac{\partial V}{\partial\Phi_L} &=&0,\nonumber \\
\dot{\Phi}_L^{2}-dH_L^{2}-V & = & 0,\nonumber \\
\dot{H}_L-H_L\dot{\Phi}_L & = & 0,\label{eq:non-local cosmo EoM}
\end{eqnarray}
where the subscribe $L$ indicates that the quantities belong to the loop corrected theory. 
For an  positive integer $n$ and arbitrary parameters $m_n$, $\sigma_n$, 
a  class of  solutions of the above EOM was constructed in refs \cite{Gasperini:2003pb, Gasperini:2004ss},

\begin{eqnarray}
\Phi^{(n)}_L(t) & = & \frac{1}{2n}\log\left(\frac{\sigma_n^{2n}}{1+\left(m_n t\right)^{2n}}\right),\\
H^{(n)}_{L} (t) & = & \frac{1}{\sqrt d} \frac{m_n}{\sigma_n} e^{\Phi^{(n)}_L \left(t\right)}= \frac{m_n}{\sqrt d} \left[\frac{1}{1+\left(m_n t\right)^{2n}}\right]^{1/2n}\nonumber.
\label{eq:nth Phi}
\end{eqnarray}
with the potential
\begin{equation}
V^{(n)}_{L}  =   \left(\frac{m_n}{\sigma_n}\right)^{2}e^{2\Phi^{(n)}_L \left(t\right)}\left[\left(1 - \sigma_n^{-2n} e^{2n\Phi^{(n)}_L \left(t\right)}\right)^{\frac{2n-1}{n}}-1\right].
\label{eq:potentials}
\end{equation}
As argued in refs \cite{Gasperini:2003pb, Gasperini:2004ss}, since
the factor $e^\Phi$ roughly represents the coupling constant, the integer $n$ effectively represents the loop number. The potential indicates  the non-perturbative contributions by the $n$th loop.

In ref. \cite{Wang:2019dcj}, a map is found between the $\alpha'$ corrected theory and loop corrected theory,

\begin{eqnarray}
H_{L} & \leftrightarrow & f\left(H_{\alpha^{\prime}}\right),\nonumber \\
-V_{L} &\leftrightarrow & g\left(H_{\alpha^{\prime}}\right)+df\left(H_{\alpha^{\prime}}\right)^{2}\nonumber \\
\Phi_L &\leftrightarrow & \Phi_{\alpha'} + \Phi_0.\label{eq:loop-alpha}
\end{eqnarray}
where $\Phi_0$ is a constant. Through this map, we thus can find a loop corrected non-perturbative non-singular solution from our $\alpha'$ corrected solution (\ref{eq:alphaSolution}).  Since  $e^{n\Phi_L^{(n)}}$ indicates the contribution from the $n$-th loop,
we use eqn. (\ref{eq:nth Phi}) to express the solution in terms of $e^{\Phi_L^{(n)}}$. Then, after some tedious calculation, we obtain the loop corrected solution,
\begin{eqnarray}
\Phi_L(t) &=& \Phi_L^{(1)} + \sum_{n=1}^\infty e^{2n\Phi_L^{(n)}} \nonumber\\
H_L(t)&=& \frac{m_1}{\sigma_1} \exp\left[\Phi_L^{(1)} +  \sum_{n=1}^\infty e^{2n\Phi_L^{(n)}}  \right] \nonumber\\
V_L(\Phi_L^{(n)}(t))&=& \left( \dot \Phi_L^{(1)} + \sum_{n=1}^\infty 2n \dot \Phi_L^{(n)} e^{2n\Phi_L^{(n)}}   \right)^2 -   \frac{d m_1^2}{\sigma_1^2} \exp\left[2\Phi_L^{(1)} + 2 \sum_{n=1}^\infty e^{2n\Phi_L^{(n)}}  \right],
\end{eqnarray}
where we set $m_n = \sqrt{2d/\alpha'}$ and $\sigma_i$'s are free parameters to be determined by coefficients calculated from the effective low energy action. Obviously, this solution is non-singular around the non-perturbative region $t\sim 0$.

\section{Conclusion}

In this paper, based on Hohm-Zwiebach action, 
we first constructed a class of general $\alpha'$ corrected non-perturbative non-singular string cosmology solutions. 
Currently, only the first two orders in $\alpha'$ correction are available. Our solution matches these results in the perturbative region as required. Once higher orders in $\alpha'$ correction are available in the future, we can straightforwardly fix the corresponding parameters in our solution to match the perturbative solutions. There exists a phenomenological map between $\alpha'$ corrected theory and loop corrected theory given in ref. \cite{Wang:2019dcj}. We used this map to constructed the corresponding  loop corrected solution, which is also non-singular and non-perturbative. Since both solutions are non-singular everywhere, the pre- and post-big-bangs are thus smoothly connected by them.

\begin{acknowledgments}
Li Song is supported by Sichuan Science and Technology Program, NO:2022YFG0317. Deyou Chen is supported by Tianfu talent plan and FXHU
\end{acknowledgments}


\begin{thebibliography}{99}
\bibitem{Gasperini:2023tus1} We are indebted to the anonymous referee to help us clarify this confusion.
\end{thebibliography}
\end{document}